\documentclass{amsart}

\usepackage{amsmath}
\usepackage{amssymb}
\usepackage{amsfonts}
\usepackage{amsthm}

\DeclareMathOperator{\diverg}{div}

\DeclareMathOperator{\im}{Im}

\DeclareMathOperator{\re}{Re}
\DeclareMathOperator{\ran}{ran}
\DeclareMathOperator{\meas}{meas}

\newtheorem{lemma}{Lemma}[section]
\newtheorem{theorem}[lemma]{Theorem}
\newtheorem{proposition}[lemma]{Proposition}
\newtheorem{assumption}[lemma]{Assumption}
\newtheorem{corollary}[lemma]{Corollary}
\theoremstyle{definition}

\theoremstyle{definition}
\newtheorem{remark}[lemma]{Remark}
\theoremstyle{definition}

{\catcode `\@=11 \global\let\AddToReset=\@addtoreset}
\AddToReset{equation}{section}

\newcommand{\N}{{\mathbb N }}
\newcommand{\R}{{\mathbb R}}
\newcommand{\C}{{\mathbb C}}
\newcommand{\e}{{\varepsilon }}
\newcommand{\ie}{{\sl i.e.\/ }}
\newcommand{\cf}{{\sl cf.\/ }}
\newcommand{\eg}{{\sl e.g.\/}}

\def\d{{\partial}}
\def\S{{\mathcal S}}
\def\v{{\tt v}}
\def\w{{\tt w}}
\def\({\left(}
\def\){\right)}
\def\<{\left\langle}
\def\>{\right\rangle}
\def\O{\mathcal O}

\newcommand{\newpar}{\par}\parindent =0pt\parskip=3pt\textheight = 615pt

\newcommand{\Id}[1]{{\rm I\kern-2pt I_{#1}}}
\renewcommand{\hbar}{{\displaystyle\bar{\phantom{x}}\kern-6pt h}}

\numberwithin{equation}{section}

\begin{document}


\title[Asymptotics for nonlinear Bloch waves]{Semiclassical
Asymptotics for Weakly Nonlinear Bloch Waves} 
\author[R. Carles]{R{\'e}mi Carles}
\author[P. A. Markowich]{Peter A. Markowich}
\author[C. Sparber]{Christof Sparber}
\address[R. Carles\footnote{On leave from MAB, Universit\'e Bordeaux
1}]{IRMAR\\ Universit\'e de Rennes 1\\ Campus de 
Beaulieu\\ 35042 Rennes\\
France} 
\email{remi.carles@math.univ-rennes1.fr}
\address[P. A. Markowich and C. Sparber]{Institut f\"ur Mathematik der
Universit\"at Wien\\ Nordbergstra\ss e 15\\ A-1090 Vienna\\ Austria}
\email{peter.markowich@univie.ac.at}
\email{christof.sparber@univie.ac.at}
\begin{abstract}
We study the simultaneous semi-classical and adiabatic
asymptotics for a class of (weakly) nonlinear Schr\"odinger equations
with a fast periodic potential and a slowly varying confinement potential. 
A rigorous two-scale WKB--analysis, locally in time, is performed. The main
nonlinear phenomenon is a modification of the Berry phase.
\end{abstract}
\thanks{This work was partially supported by the EU network
HYKE (contract no. HPRN-CT-2002-00282), the Wittgenstein Award 2000 of 
P. A. M. (funded by the Austrian research fund FWF), and the
Wissenschaftskolleg  Differentialgleichungen (FWF project no. W8).}
\maketitle



\section{Introduction and scaling}

In this work we study the asymptotic behavior as $\e \rightarrow 0$ of
the following semilinear initial value problem (IVP):
\begin{equation}
\label{nls}
\left \{
\begin{aligned}
i\e \partial _t \psi^\e = & -\frac{\e^2}{2}\Delta \psi^\e +
V_{\Gamma}\left(\frac{x}{\e}\right)\psi^\e + U(x)\psi^\e + 
\e \lambda(t)\, |\psi^\e|^{2\sigma} \psi^\e,\\
\psi^\e \big |_{t=0}   = &  \ \psi^\e_I(x),
\end{aligned}
\right.
\end{equation}
where $x \in \R^d$, $t\in \R $, $ \sigma \in \N$ and $0< \e \ll1$. 
Here and in the following $\e$-dependence will be denoted by the 
superscript $\e$. The external (confining) potential $U=U(x)\in \R$ is
assumed to be smooth on $\mathbb R^d$, whereas the lattice-potential
$V_\Gamma= V_\Gamma(y)\in \R$ is assumed to be smooth, uniformly bounded in
$\R^d$ and \emph{periodic} with respect to some \emph{regular lattice}
$\Gamma \simeq \mathbb Z^d$, generated through a basis
$\{\zeta_1,\dots,\zeta_d\}$, $\zeta_l \in \R^d$, \ie  
\begin{equation}
\label{eq:Vper}
V_{\Gamma}(y + \gamma) = V_{\Gamma}(y), \quad \forall y \in \R^d,
\gamma \in \Gamma, 
\end{equation}
where
\begin{equation}
\label{eq:net}
\Gamma=\left\{\gamma \in \R^d: \ \gamma=\sum_{l=1}^d \gamma_l \zeta_l, 
\ \gamma_l\in \mathbb Z \right\}.
\end{equation}
Finally, we assume $\lambda=\lambda (t)\in \R $ to be a smooth
coupling-function and $\psi_I^\e\in L^2(\R^d)$ to be normalized such that  
\begin{equation}
\label{nor}
\int_{\R^d} |\psi_I^\e(x) |^2 dx =1.
\end{equation}
This normalization is henceforth preserved by the evolution since
$\lambda(t)\in \R$.    
\newpar 
Nonlinear Schr\"odinger equations (NLS) of type \eqref{nls} appear in
various physical situations, \cf \cite{SuSu} for a general
overview. An important example in $d=3$ is the case $\sigma =1$,  
$\lambda(t)\equiv \pm 1$, \ie  the so called \emph{repulsive} resp. 
\emph{attractive Gross--Pitaevskii equation}, a celebrated model for
the description of the evolution of \emph{Bose--Einstein condensates}
(BECs) \cite{PiSt}. In order to motivate the scaling in \eqref{nls} we
shall examine this case more closely: 
\newpar
In physical units, the Gross--Pitaevskii equation (for $d=3$) is given
by \cite{PiSt} 
\begin{equation}
\label{gpe}
i\hbar \partial _t \psi =  -\frac{\hbar^2}{2m}\Delta \psi +
V\left(x\right)\psi + U_0(x)\psi \pm 
N\alpha(t) |\psi|^{2} \psi,
\end{equation}
where $m$ is the atomic mass, $\hbar$ is the Planck constant, 
$N$ is the number of atoms in the condensate and 
\begin{equation} 
\alpha(t) = \frac{4\pi\hbar^2|a(t)|}{m},
\end{equation}
with $a(t) \in \R$ denoting the $s$-wave scattering length derived from the 
corresponding $N$-particle theory, \cf \cite{LSY, PiSt}. (The fact that $a(t)$ is 
chosen time-dependent is motivated by recent experiments on BEC where this has 
indeed be achieved by some highly sophisticated experimental techniques.) In this
context the external potential $U(x)$, which traps the condensate, is
usually assumed to be a harmonic confinement potential of the  
following form \cite{BJM, DFK}:
\begin{equation}
U_0(x) = \frac{m\omega_0^2}{2} \, |x|^2, \quad \omega_0\in \R, \, x\in \R^3.
\end{equation}
More general, non-isotropic variants of such confinement potentials
are used to create so called \emph{disc-shaped} or
\emph{cigar-shaped}, \ie quasi two or, resp., one dimensional,  
BECs (see \cite{BJM, PiSt} and the references given therein). 
If in addition a periodic potential $V(x)$, which in physical
experiments is generated by an intense laser field, is included, the
condensates are referred to as \emph{lattice BECs}. A particular
example of $V$ is then given by  
\begin{equation}
\label{lasp} 
V\left(x\right)= \sum_{l=1}^3 \frac{\hbar^2 \xi_l^2}{2m} 
\sin^2\left(\xi_{l} x_l \right),
\end{equation}
where $\xi=(\xi_1, \xi_2,\xi_3)$ with $\xi_l\in \R$ denotes the wave vector of the laser field \cite{PiSt}. 
The sign in front of the nonlinearity in \eqref{gpe} corresponds to a \emph{stable} (defocusing) resp. 
\emph{unstable} (focusing) condensate. To rewrite the equation \eqref{gpe} into our semi-classical scaling we
proceed similar to \cite{BJM}. More precisely, we introduce
dimensionless variables 
\begin{align}
\tilde t = \omega_0\, t, \qquad \tilde x =\frac{x}{x_s},\qquad
\tilde\psi(\tilde t, \tilde x) = x_s^{3/2} \psi(t,x), 
\end{align}
where $x_s$ will be determined later and $\tilde \psi(\tilde t, \tilde x)$ 
is such that the normalization \eqref{nor} is preserved 
for $d=3$. Multiplying \eqref{gpe} by $1/(m \omega_0^2 x_s^2)$ and
omitting again all "$ \ \tilde { } \ $" we find the following
dimensionless equation: 
\begin{equation}
i\e \partial _t \psi =  -\frac{\e^2}{2}\Delta \psi +
V_{\Gamma}\left(\frac{x}{\varepsilon}\right)\psi + U(x)\psi \pm 
\delta(t) \e^{5/2} |\psi|^{2} \psi,
\end{equation}
where the potentials are defined by 
\begin{equation}
\label{pode}
V_{\Gamma}\left(y \right):= \frac{V( x_s \e y)}{m \omega_0^2 x_s^2}\, , \qquad 
U(x):= \frac{|x|^2}{2}\, , 
\end{equation}
and the appearing parameters $\e$, $\delta(t) \in \R_+$ are
\begin{equation}
\label{par}
\e := \frac{\hbar}{\omega_0 m x_s^2}=\left(\frac{a_0}{x_s}\right)^2, \qquad 
\delta(t) := \frac{N\alpha(t)}{a_0^3 \hbar \omega_0} =
\frac{4\pi |a(t)| N}{a_0}\,,
\end{equation}
with $a_0$ denoting the length of the harmonic oscillator ground state
corresponding to $U_0(x)$, \ie 
\begin{equation}
a_0:= \sqrt{\frac{\hbar}{\omega_0 m}}.
\end{equation}
Since we aim for $\e\ll 1$ and $\delta \e^{5/2}$ to be of the order of
$\e$ we require  $\delta = O(\e^{-3/2})$, hence $4\pi |a| N \gg a_0$,
which from a physical point of view corresponds to the \emph{strong
interaction regime}, also known as \emph{Thomas--Fermi regime}
\cite{PiSt}. Now, consider a reference value $\bar a$ for $a(t)$  
and similarly denote by $\bar \delta$ the parameter $\delta $ for this
reference value $\bar a$. Inserting \eqref{par} into $\bar \delta
\e^{5/2} = \e$, we compute the \emph{characteristic length scale}  
\begin{equation}
\label{xs}
x_s = ( 4\pi N |\bar a| a_0^2 )^{1/3},
\end{equation}
which one needs to choose as the appropriate reference scale in our
situation. In particular we shall assume $|\psi^\e_I(x)|$ to vary on
this scale. The coupling function $\lambda(t)$ is then given by  
$\lambda (t) = \delta(t)/\bar \delta$. Identity \eqref{xs} implies 
\begin{equation}
\e = \left ( \frac{a_0}{4\pi N |\bar a|}\right)^{2/3}\ll 1,
\end{equation}
which is different from the one given in \cite{BJM}. Moreover, having
in mind \eqref{lasp}, \eqref{pode} we require for the periodic
potential $V_\Gamma$  
\begin{equation}
\label{conp}
\e \xi_l x_s  = O(1), \qquad \frac{\hbar^2 \xi_l^2}{2m^2 x_s^2
\omega_0^2} = O(1), \quad \mbox{for $l=1,2,3$}. 
\end{equation}
From these relations one computes 
\begin{equation}
\xi_l \approx a_0^{-4/3}(4\pi N |\bar a|)^{1/3},\quad \mbox{for $l=1,2,3$},
\end{equation}
which gives the required wave vector in our regime and one checks that
in this case the conditions \eqref{conp} are satisfied. We remark that
this scaling is in good agreement with some typical recent
experiments. For example in the case of a lattice BEC consisting of Rb
atoms we have, \cf \cite{BJM, DFK}: 
\begin{equation}
a_0\approx 3,4 \times 10^{-6} [m], \quad \bar a\approx 5,4 \times
10^{-9} [m], \quad N\approx 1,5 \times 10^{5}. 
\end{equation}
This gives: $4\pi |\bar a| N \approx 10^{-2} [m] \gg a_0$, hence $\e
\approx 4,3\times 10^{-3}\ll1 $ and for the wave vectors we compute
$\xi_l \approx 4,6\times 10^{6} [1/m]$, which is of the same
order of magnitude as stated in \cite{ChNi}. The reference length
scale in this case is $x_s=2,1\times 10^{-6} [m]$, which is $O(a_0)$. 
Finally, to motivate the choice $\sigma \geq 1$, we
note that for $d<3$ higher order nonlinearities are frequently used in
the description of BECs \cite{KNSQ, LSY}.  
\newpar
From a mathematical point of view the limit $\e \rightarrow 0$
corresponds to the simultaneous \emph{semi-classical} (or
\emph{high-frequency}) and \emph{adiabatic limit} (see \cite{MaSp, Ro,
Te} for general introductions to these fields). For \emph{linear}
time-dependent Schr\"odinger equations (with periodic potentials) this
asymptotic regime has been intensively studied by several authors,
using (spatial)  \emph{adiabatic decoupling theory} \cite{PST, Te} 
or \emph{Wigner measures} \cite{BFPR, Ge, GMMP}, to mention
results obtained in recent years. A numerical study
of these asymptotics can be found in \cite{GoMa}. 
\newpar 
In our scaling the nonlinearity is $o(1)$ and can thus be
called \emph{weak}, still it makes the rigorous asymptotic analysis of
the given IVP considerably harder. Even without a periodic potential
the semi-classical limit for NLS is still far from being completely
understood.  
In particular, we cannot use the above mentioned mathematical techniques, 
which so far only work in a linear setting. (For a notable exception
see \cite{BMP}.)  
Thus we shall rather apply a more naive asymptotic expansion method in the 
spirit of the traditional \emph{WKB--type expansions}. 
Due to the periodic potential, we use a so called \emph{two-scale WKB--ansatz}, 
first introduced in \cite{BLP}, which has already been successfully
applied in the case of linear  
periodic Schr\"odinger equations \cite{DGR, GRT}. 
Our scaling is such that the nonlinearity enters in the leading order
term of the asymptotic WKB--type solutions, although the
Hamilton-Jacobi equation for the phase of the wave--function is found to be the same as in the linear
case.  
This is analogous to the \emph{weakly nonlinear (dispersive)
geometrical optics} regime discussed  
in \cite{DoRa}. 
(See also \cite{SpMa} for an application of this scaling in another 
semi-classical context). 
The asymptotic description is valid on \emph{macroscopic} time-scales
$t=O(1)$ but in general only for small $|t|>0$. 
 
Before giving a precise description, we state the typical result that
we shall prove. The possibly not well-defined assumptions in the
following statement will be discussed more precisely below.

\begin{theorem}\label{theo:typique}
Let $d\geq 1$, $V_\Gamma$ and $U$ be smooth, real-valued potentials, 
$V_\Gamma$ being $\Gamma$-periodic, $U$ being sub-quadratic, and
$\lambda$ being real-valued and smooth.  
Assume that the initial datum $\psi^\e_I$ is of the form 
\begin{equation*}
\psi^\e_I (x) = a_I(x)\chi_n\(\frac{x}{\e},\nabla
\phi_I(x)\)e^{i\phi_I(x)/\e} + \e \varphi_I^\e(x),
\end{equation*}
where $a_I\in \S(\R^d;\C)$, $\phi_I\in C^\infty(\R^d;\R)$ and
$\chi_n=\chi_n(y,k)$ is a Bloch eigenfunction associated to a simple isolated
Bloch band $E_n=E_n(k)$. We suppose that  $\varphi_I^\e$ satisfies
Assumption~\ref{ass:varphi} below, with $K\ge d$.
Assume that no caustic is formed before time
$\tau>0$, and fix $\tau_0\in]0,\tau[$.  
Then there exists $\e_0>0$ such that for $0<\e\leq
\e_0$, the solution $\psi^\e$ to \eqref{nls} is defined up to time
$\tau_0$. Moreover, it satisfies the following asymptotics as $\e\to
0$: 
\begin{align}
\sup_{0\leq t\leq \tau_0}\left\| \psi^\e(t) -
\v_0^\e(t)
\right\|_{L^2(\R^d)}&=\O(\e) ,\nonumber \\ 
\sup_{0\leq t\leq \tau_0}\left\| \psi^\e(t) -
\v_0^\e(t)
\right\|_{L^\infty(\R^d)}&=\O\(\e^{1-\eta}\) ,\quad \text{for any
}\eta>0\, ,\label{eq:asymLinfty}
\end{align}
where the approximate solution $\v_0^\e$ is given by:
\begin{equation*}
\v_0^\e(t,x)= \frac{a_I
\(X^{-1}_t(x)\)}{\sqrt{J_t\(X^{-1}_t(x)\)}}\chi_n\(  
\frac{x}{\e},\nabla_x\phi(t,x)\) e^{i\omega\(t,X_t^{-1}(x)\) } 
e^{i\phi(t,x)/\e}\, .
\end{equation*}
Here, $\phi$ solves the Hamilton-Jacobi equation \eqref{hj}, corresponding 
to the classical flow: $(t,x)\mapsto X_t(x)$, as defined by \eqref{semi}, $J_t$
is the associated Jacobi determinant \eqref{eq:determinant}, and 
$\omega$ is given by
\begin{align*}
\omega(t,x)=& \, -i\int _0^t\beta\(s,X_s(x) \) ds \\
&-|a_I(x)|^{2\sigma} \int_0^t
\frac{\lambda(s)}{J_s(x)^{\sigma}} \int_Y \left|\chi_n\(
y,\nabla_x\phi\(s,X_s(x)\)\) \right|^{2\sigma+2}dy\, ds.
\end{align*}
We denote by 
$\beta \in i\R$ the Berry phase \eqref{b},
and by $Y$ the centered fundamental domain of $\Gamma$. 
\end{theorem}
\begin{remark}
Our result holds only before caustics. This should not be surprising;
even in the linear case $\lambda\equiv 0$, the WKB method is
effective only away from caustics. On the other hand, some techniques
have proved to be efficient to overcome this difficulty in a linear
framework, such as Gaussian beams (see e.g. \cite{DGR}) or Wigner
functions (see e.g. \cite{LionsPaul,SpMaMa}). However, adapting these
techniques to a 
nonlinear context seems to be a challenging open question. 
\end{remark}
\begin{remark}
The assumptions on the corrector $\varphi_I^\e$ for the initial data
are not trivial (see Assumption~\ref{ass:varphi}). They state
essentially that the initial data are well-prepared, in order to prove
a nonlinear stability result. Note however that $\varphi_I^\e$ is of order
$\O(1)$ as $\e\to 0$ in any reasonable sense. The assumptions $K\ge d$
means that we have to consider (at least) $d$ correctors to prepare the
initial data. This assumption may seem surprising; the proofs we give
rely on it, and it would be interesting to understand how necessary
this assumption is.  
\end{remark}
The above result shows that the leading order nonlinear phenomenon is
represented by the phase factor $\omega$. The Berry phase is a linear
(geometrical) feature (see \eqref{b} below), but the second 
integral in the definition of $\omega$ stems from the nonlinearity. In
the context of laser physics, this phenomenon is known as
\emph{phase self-modulation} (see e.g. \cite{ZS,Boyd,Donnat}). 

The paper is organized as follows. In Section~\ref{sec:DA.1}, we start
a formal asymptotic expansion, following WKB--methods. This leads us to
consider the Bloch eigenvalue problem. The asymptotic expansion is
considered in more detail in Section~\ref{sec:DA.2}, where a
formal approximate solution is constructed at any order. The
justification of this approximation is performed in
Section~\ref{sec:stab}. We discuss our results and some of their
possible extensions in Section~\ref{sec:disc}. In
Appendix~\ref{sec:appA}, we detail a computational step from
Section~\ref{sec:DA.2}.  
 

\section{Asymptotic expansion: emergence of Bloch bands}\label{sec:DA.1}

For solutions of \eqref{nls} we seek an asymptotic expansion of
the following form:  
\begin{equation}
\label{wkb}
\psi^\e (t,x)  =   \ u^\e \left(t,x,
\frac{x}{\e}\right)e^{i\phi(t,x)/\e}\quad ; \quad
u^\e (t,x,y) \sim  \ \sum_{j=0}^\infty \e^j u_j(t,x,y), 
\end{equation}
where we assume that both $\phi(t,x)\in \R$ and $u^\e(t,x,y)\in \C$ 
are sufficiently smooth. Moreover we impose
\begin{equation*}
u^\e(\cdot,\cdot,y + \gamma) = u^\e(\cdot,\cdot,y), \quad \forall \, y
\in \R^d,  
\, \gamma \in \Gamma.
\end{equation*}
We assume that the initial condition $\psi^\e_I$ is 
compatible with \eqref{wkb}:
\begin{assumption} The initial wave--function $\psi_I^\e$ is in the
Schwartz space
$\mathcal S(\R^d)$, and is of WKB--type, \ie
\label{assi}
\begin{equation}
\psi_I^\e(x)= u_I\left(x,\frac{x}{\e}\right)e^{i\phi_I(x)/\e} +\e
\varphi_I^\e(x), 
\end{equation} 
with $\phi_I \in C^\infty(\R^d;\R)$, $u_I\in {\mathcal S}(\R^d\times \mathbb
T^d; \C)$\footnote{That is, $u_I$ is rapidly decaying w.r.t. the first
variable ($x$), 
smooth w.r.t. the second one ($y$).}, $\mathbb T^d\equiv
\R^d/  \Gamma$. The function $\varphi_I^\e$ is a corrector to be
precised later on.  
\end{assumption}
From now on we shall denote the linear part of the Hamiltonian
operator by  
\begin{equation}
\label{ham}
H^\e: =  - \frac{\e^2}{2} \Delta +
V_{\Gamma}\left(\frac{x}{\e}\right) + U(x)
\end{equation}
Plugging the ansatz \eqref{wkb} into \eqref{nls} we (formally) obtain: 
\begin{equation*}
i\e \partial _t \psi^\e - H^\e \psi^\e- 
\e \lambda (t)  |\psi^\e|^{2\sigma}\psi^\e  =
b^\e \(t,x,\frac{x}{\e}\) e^{i\phi(t,x)/\e} .\\
\end{equation*}
We consequently expand the r.h.s. of this equation as  
\begin{equation}
\label{exp}
b^\e(t,x,y)\sim  \sum_{j=0}^\infty \e^j b_j(t,x,y) 
\end{equation}
and choose the asymptotic amplitudes $u_j$ in a way such that
$b_j(t,x,y)\equiv 0$, $\forall j\geq0$.  
\newpar
Setting $b_0(t,x,\frac{x}{\e})=0$ yields
\begin{align}
\label{r0.0}
-\frac{\Delta_y u_0}{2} - i\nabla_x\phi\cdot \nabla_y u_0 +
 \frac{|\nabla_x\phi|^2}{2}u_0+V_\Gamma(y)u_0
 +\left(U(x)+\partial_t \phi  \right) u_0\big|_{y=\frac{x}{\e}}=0. 
\end{align}
Uncorrelating the variables $x$ and $y$, we shall seek a solution to
the more general equation: 
\begin{align}
\label{r0}
-\frac{\Delta_y u_0}{2} - i\nabla_x\phi\cdot \nabla_y u_0 +
 \frac{|\nabla_x\phi|^2}{2}u_0+V_\Gamma(y)u_0 = -\left(U(x)+\partial_t \phi
 \right) u_0\, .
\end{align}
Denoting by 
\begin{equation}
\label{hamp}
H_{\Gamma}(k): = \frac{1}{2} \, \left(-i\nabla_y + k \right)^2+
V_{\Gamma}\left (y\right), 
\quad k\in \R^d,
\end{equation}
we can rewrite equation \eqref{r0} in the following form:
\begin{align}
\label{r00}
H_{\Gamma} (\nabla_x \phi)u_0 = -\left(U(x)+\partial_t \phi  \right) u_0.
\end{align}
We now require that for some fixed $n\in \N$, it holds 
\begin{equation}
\label{hj}
E_n(\nabla_x \phi)=-\left(U(x)+\partial_t \phi  \right) ,
\end{equation}
where $E_n(k)$, $k\in \R^d$, is the $n$-th eigenvalue of the
\emph{Bloch eigenvalue problem} \cite{Bl}: 
\begin{equation}
\label{bloch}
\left \{
\begin{aligned}
H_{\Gamma}(k) \chi_n(y,k) = & \ E_n(k)\chi_n(y,k),\quad n\in \N,\, y\in Y,\\
\chi_n(y+\gamma,k)=& \ \chi_n(y,k), \quad \mbox{for $\gamma \in \Gamma$}.
\end{aligned}
\right.
\end{equation}
Here and in the following, we denote by $Y$ the centered
\emph{fundamental domain}  of the lattice $\Gamma$, \ie
\begin{equation}\label{eq:Y}
Y:= \left\{\gamma \in \R^d: \ \gamma=\sum_{l=1}^d \gamma_l \zeta_l, 
\ \gamma_l\in \left[-\frac{1}{2}, \, \frac{1}{2}\right] \right\},
\end{equation}
whereas $Y^*$, denotes the corresponding basic cell of the dual
lattice $\Gamma^*$.  
In solid state physics $Y^*$ is called the \emph{Brillouin zone} hence we 
shall denote it by $\mathcal B\equiv Y^*$. 
Let us recall some well known facts for this eigenvalue problem, \cf
\cite{Ne, Te, Wi}: 
\newpar 
Since $V_\Gamma$ is smooth and periodic, we get that, 
for every fixed $k\in \mathcal B$, $H_\Gamma(k)$ is self-adjoint on
$H^2(\mathbb T^d)$ with compact resolvent.  
Hence the spectrum of $H_\Gamma(k)$ is given by
\begin{equation*}
\sigma (H_\Gamma(k))= \{
E_n(k)\ ;\ n\in \N^*\}, \quad E_n(k)\in \R. 
\end{equation*}
In general we can order the eigenvalues $E_n(k)$ according to their
magnitude and multiplicity, 
$$
E_1(k)\leq\ldots\leq E_n(k)\leq E_{n+1}(k)\leq \dots
$$
Moreover every $E_n(k)$ is periodic w.r.t. $\Gamma^*$ and it holds that 
$E_n(k)=E_n(-k)$. The set $\{E_n (k); \, k \in \mathcal B\}$ is called the 
$n$th-\emph{energy band}.
The associated eigenfunction, the \emph{Bloch waves}, $\chi_n(y,k)$ 
form (for every fixed $k\in\mathcal B$) a complete orthonormal basis
in $L^2(Y)$  
and are smooth w.r.t. $y\in Y$. We choose the usual normalization  
\begin{equation}
\label{norm}
\left <\chi_n(\cdot, k), \chi_m(\cdot, k) \right>_{L^2(Y)}\equiv 
\int_Y \overline{\chi_n(y, k)}\chi_m(y, k)dy =
\delta_{n,m},\quad n,\,m\in\N. 
\end{equation} 
Concerning the dependence on $k\in \mathcal B$, it has been shown
\cite{Ne} that for any $n\in \N$ there exists a closed subset
$\mathcal U\subset \mathcal B$  
such that: $E_n(k)$ are analytic, $\chi_n(\cdot, k)$ can be chosen to
be analytic functions   
for all $k \in \Omega:=\mathcal B \backslash \mathcal U$, and 
\begin{equation}
\label{iso}
E_{n-1} < E_n(k) < E_{n+1}(k),\quad \forall k \in \Omega.
\end{equation}
If this condition holds for all $k\in \mathcal B$ then $E_n(k)$ is
called an \emph{isolated  
Bloch band} \cite{Te}. Moreover, it is known that 
$$
\meas \mathcal U = \meas \, \{ k\in\mathcal B\ | \ E_n(k)=E_{m}(k), \ n\not = m \}=0.
$$ 
In this set of measure zero one encounters so called \emph{band crossings}. 
\newpar
Equation \eqref{hj} is called the $n$-th band \emph{Hamilton-Jacobi equation} 
corresponding to the \emph{semi-classical band Hamiltonian}
\begin{equation}
\label{sch}
h^{sc}_n(k,x):=E_n(k)+ U(x),\quad (k,x)\in \mathbb T^*\times \R^d, 
\end{equation} 
with an effective kinetic energy given by the $n$-th eigenvalue for 
$k\in \mathbb T^*\equiv\R^d/ \Gamma^*$. 
The characteristic differential equations corresponding to \eqref{hj}
are consequently given by the equations of motion:  
\begin{equation}
\label{semi}
\left \{
\begin{aligned}
\dot x = & \ \nabla_k E_n(k),\quad x\big|_{t=0}=x_0\in \R^d,\\
\dot k = & \ -\nabla_x U(x) ,\quad k\big|_{t=0}=\nabla_x \phi_I(x_0).
\end{aligned}
\right.
\end{equation}
This system (locally) defines a flow map $(x,t)\mapsto
X_t(x)\equiv X_t(x;\nabla_x\phi_I(x))$ in physical space. In general
\emph{caustics} will appear in this flow, which prohibits the
existence of globally defined smooth solutions for \eqref{hj}. Let us
denote by 
\begin{equation}\label{eq:determinant} 
J_t(x):=\det\( \nabla_{x} X_t(x;\nabla_x\phi_I(x))\)
\end{equation}
the corresponding Jacobi determinant. We have $J_0(x)\equiv 1$. Denote  
by $\tau$ the time at which the first caustic appears, \ie 
\begin{equation}\label{eq:caustique}
\tau := \inf\{t>0 \ | \ \exists \, x\in \mathbb R^d:  J_t(x)=0\}.
\end{equation}
We thus have $J_t(x)>0$ for $0\leq t<\tau$. 
Standard theory implies the following:
\begin{lemma} \label{le0}
If $h^{sc}_n(k,x)\in C^\infty(\mathbb T^*\times\R^d)$, $\phi_I\in
C^\infty(\mathbb R^d)$, then there  
exist  $\tau >0$ and a unique smooth solution $\phi \in C^\infty ([0,
\tau[ \times \R^d )$ of the Hamilton-Jacobi equation 
\begin{equation*}
\partial_t \phi + h^{sc}_n(\nabla_x \phi,x)=  \ 0\quad ; \quad
\phi\big|_{t=0}=  \ \phi_I(x).
\end{equation*}
\end{lemma} 
To make sure that $E_n(k)$ (and hence $h^{sc}_n(k,x)$) 
is sufficiently smooth, we shall impose the following assumption: 
\begin{assumption}
\label{ass}
The amplitude $u_I(x,y)$ is assumed to be concentrated in a single
isolated Bloch band $E_n(k)$ corresponding to a simple
eigenvalue of $H_\Gamma(k)$, \ie  
\begin{equation}
\label{u}
u_I(x,y)\equiv a_I(x) \chi_n(y,\nabla_x \phi_I(x)),
\end{equation}
where $a_I\in \mathcal S(\R^d; \C)$ is a given initial
amplitude. 
\end{assumption}
From~\eqref{r00} and~\eqref{bloch} we conclude that there exists
$a_0=a_0(t,x)$ such that  
\begin{equation}
u_0(t,x,y) = a_0(t,x) \chi_n(y, \nabla_x\phi(t,x)).
\end{equation}
\begin{remark} 
Note that also in the linear case, assumptions similar to
Assumption~\ref{ass} are usually imposed, \cf \cite{GMMP, PST}. There
however, the reason is largely to avoid band crossings in order to
obtain global-in-time results. (The rigorous study of band crossings
is quite involved and up to now established only for certain model
problems, \cf \cite{FeGe, FeLa, Ha}.)\\  
Due to caustics (and possibly additional nonlinear effects if
$\lambda(t)$ is not 
real-valued, see Sect.~\ref{sec:disc}), we cannot hope for such 
global-in-time results in our case. Assumption~\ref{ass} therefore is
only imposed for regularity reasons and could be significantly
weakened, since, with some technical effort, one could modify the
subsequent analysis. Indeed, all statements could be formulated locally in
regions $\mathcal U\subseteq \R_t\times \R^d_x$ which neither contain
caustics nor band crossings (in the sense that $E_n(\nabla_x
\phi(t,x))\not = E_m(\nabla_x \phi(t,x))$, for all $(t,x)\in \mathcal
U$). In this way one could include also non-isolated bands $E_n(k)$.\\ 
We further remark that in the case $d=1$ all band crossings can be
removed through a proper analytic continuation of the bands, \cf
\cite{ReSi}.   
\end{remark}


\section{Derivation of the transport equations}\label{sec:DA.2}

To characterize the \emph{principal amplitude} $a_0$,  we set $b_1=0$
in \eqref{exp}, which yields 
\begin{align}
\label{r01}
H_{\Gamma} (\nabla_x \phi)u_1+\left(U(x)+\partial_t \phi  \right) u_1=  
L_1 u_0 - \lambda(t) |u_0|^{2\sigma}u_0,
\end{align}
where the linear differential operator $L_1$ applied to $u_0$ reads
\begin{equation}
\label{l1}
L_1u_0 :=i \partial_t u_0 + i\nabla_x \phi\cdot\nabla_x u_0 +
i\frac{\Delta_x \phi}{2} u_0 
+\diverg_x\nabla_yu_0.
\end{equation}
We multiply equation \eqref{r01} with $\overline
\chi_n(y,\nabla_x\phi)$ and integrate over  
the fundamental domain $Y$. From \eqref{hj}, the left hand side of
\eqref{r01}  is $(H_\Gamma -E_n)u_1$; since $H_{\Gamma}$ is 
self-adjoint, the integral
obtained from the 
l.h.s. of \eqref{r01}  
is identically zero, hence:
\begin{equation}
\label{cond}
\int_{Y} \overline \chi_n(y,\nabla_x\phi) \, \left( L_1 u_0 -
\lambda(t) |u_0|^{2\sigma}u_0 \right) \, dy = 0, 
\end{equation}
is a necessary condition such 
that \eqref{r01} can be solved for $u_1$ in terms of $u_0$. This
condition is known to be sufficient, from the orthogonal decomposition method
(also known as ``Feschbach method''), since $E_n$ is an isolated eigenvalue.
After some lengthy computations, given in the appendix, we find that
\eqref{cond} is equivalent to the following \emph{nonlinear transport
equation} for $a_0$: 
\begin{equation}
\label{trans}
\left \{
\begin{aligned}
\partial_t a_0 + \mathcal L a_0 - \beta(t,x) a_0=& \ i\kappa(t,x)
|a_0|^{2\sigma} a_0,\\ 
a_0\big|_{t=0}= & \ a_{I}(x).
\end{aligned}
\right.
\end{equation}
Here, $\mathcal L$ is the usual (geometrical optics) transport operator 
associated to $h_n^{sc}(k,x)$:
\begin{equation}
\label{l}
\mathcal L a_0:= \nabla_k E_n(\nabla_x\phi)\cdot\nabla_x a_0
+ \frac{1}{2} \diverg_x(\nabla_k E_n(\nabla_x\phi))a_0.
\end{equation}
Moreover, we have 
\begin{equation}
\label{b}
\begin{aligned}
\beta (t,x):= & \, \left <\chi_n(\cdot,\nabla_x\phi),\, 
\nabla_k \chi_n(\cdot, \nabla_x\phi)\right>_{L^2(Y)}\cdot \nabla_x U(x)\\
\equiv & \, \sum_{l=1}^d  \big <\chi_n(\cdot, \nabla_x\phi), \, 
\frac{\partial }{\partial k_l} \chi_n(\cdot, \nabla_x\phi)
\big>_{L^2(Y)}\, \frac{\partial}{\partial x_l}\, U(x)  
\end{aligned}
\end{equation}
and 
\begin{equation}
\label{k}
\kappa(t,x) :=  -\lambda(t) \int_{Y}
\left|\chi_n\(y,\nabla_x\phi(t,x)\)\right|^{2\sigma 
+2} \, dy .   
\end{equation}
This term can be interpreted as an \emph{effective coupling} of the 
self-interaction within the $n$th-energy band.
\newpar
Note that \eqref{norm} implies 
$$
\re \left <\chi_n(\cdot, k),\,  \nabla_k \chi_n(\cdot, k)
\right>_{L^2(Y)} \equiv 0. 
$$
Hence, $\beta(t,x) =i \im \beta(t,x)$ only contributes a variation in
the phase of $a_0$,  
the so called \emph{Berry phase} \cite{ShWi, Te}. It is due to the
interaction of the lattice and the  
slowly varying potential $U$.  
In our case the Berry phase in addition gets modulated in a
\emph{nonlinear} way  
by the right hand side of \eqref{trans}. 

\begin{remark}
The term $-i \beta =: \mathcal A_n$ can be interpreted as a gauge potential, 
\ie a connection in the (complex) eigenspace-bundle corresponding to 
$E_n(k)$, \cf \cite{Te}. For some particular lattice 
configurations (if the crystal has a center of inversion, or some
other special symmetry), 
the curvature of the Berry connection $\Omega_n:=\nabla\times\mathcal A_n$ is 
identically zero, and the Berry connection is a closed $1$-form, 
\cf \cite{PaTe, ShWi, Te} for a broader discussion on this. 
\end{remark}
\begin{remark}
We provide a link with some already existing results. In 
\cite{PST, Te} the authors, roughly speaking, prove that 
in each isolated Bloch band $E_n(k)$ the linear Hamiltonian 
$H^\e$, defined in \eqref{ham}, can be unitarily mapped 
into an \emph{effective} band Hamiltonian $h_n^\e$, which is 
the Weyl quantization of the semi-classical symbol 
\begin{equation*} 
h_n^\e(k,x)\sim h_n^{sc}(k,x) + \e h_1 (k,x)+ O(\e^2).
\end{equation*} 
This is done by constructing an $\e$-dependent unitary operator, 
which block-diagonalizes the \emph{Bloch-Floquet Hamiltonian} of the system, 
such that the relevant band decouples from the rest of the spectrum. 
Above the \emph{principal symbol} $h_n^{sc}(k,x)$ is defined as in 
\eqref{sch} and the first order correction is such that 
\begin{equation*} 
h_1(\nabla_x \phi(t,x) , x)\equiv i\beta (t,x).
\end{equation*} 
Additional terms appear in $h_1(k,x)$ if one includes external
magnetic fields too, \cf \cite{PST,Te}. 
\end{remark}

The following lemma proves that \eqref{trans} has a smooth solution up
to caustics: 
\begin{lemma} 
\label{le1}
Assume $\phi\in C^\infty([0,\tau[\times\R^d)$, and $a_{I}\in
\mathcal S(\R^d;\C)$.  
Then along the  flow $(t,x)\mapsto X_t(x)$, \eqref{trans} has a unique
solution $a_{0}\in 
C^\infty([0,\tau[; \mathcal S(\R^d))$, given by:
\begin{equation*}
a_0(t,X_t(x))=\frac{a_{I}(x)}{\sqrt{J_t\(x\)}}\, \exp
\(i |a_I(x)|^{2\sigma} \int_0^t
\frac{\kappa \(s,X_s(x) \)}{|J_s\(x \)|^{\sigma}}\, 
ds +  \int _0^t\beta\(s, X_s(x) \) \, ds\).
\end{equation*} 
\end{lemma}
\begin{proof} 
Using Liouville's formula, 
\begin{equation*}
\frac{d}{dt}\, J_t (x)=  \ \diverg_x\Big(\nabla_k E_n\big(\nabla_x\phi
\(t,X_t(x)\)\big)\Big)\, J_t(x)\quad ; \quad 
J_{0}(x)=  \ 1\, ,
\end{equation*}
we rewrite the transport equation \eqref{trans} as an 
ordinary differential equation along the flow defined by the
dynamical system \eqref{semi}. Let $\alpha_0(t,x):=a_0(t,X_t)$: 
\begin{equation*}
\frac{1}{\sqrt{J_t(x)}} \frac{d}{dt}\, (\sqrt{J_t(x)} \alpha_0) =
\beta\(t,X_t\) \alpha_0+ 
i\kappa\(t,X_t\) |\alpha_0|^{2\sigma} \alpha_0\ ,\quad |t|<\tau.
\end{equation*}
If we define $\tilde \alpha_0:=\sqrt{J_t(x)}\alpha_0$, then
the principal amplitude is determined by
\begin{equation}
\label{eq:transray}
\left \{
\begin{aligned} 
\frac{d}{dt}\, \tilde \alpha_0 = &\beta\(t,X_t(x)\) \tilde \alpha_0+ \
i\kappa\(t,X_t(x)\) 
\frac {|\tilde \alpha_0|^{2\sigma}}{|J_t(x)|^\sigma}\, \tilde \alpha_0,\quad
|t|<\tau, \\  
\tilde \alpha_0\big|_{t=0}= & \ a_{I}(x).
\end{aligned}
\right.
\end{equation}
This implies (since $\beta(t,x)\in i\R$ and $\kappa(t,x)\in \R$)
\begin{equation*}
\frac{d}{dt}\, |\tilde \alpha_0(t,x)|^2 = 0\, ,\quad\text{hence
}|\tilde \alpha_0(t,x)|\equiv |a_I(x)| \, ,\quad \forall \,
t\in[0,\tau[ \, .
\end{equation*}
Define the phase shift $g$ of $\tilde \alpha_0$ by
$\tilde\alpha _0(t,x)=a_I(x)e^{ig(t,x)}$. Then $g$ solves 
$$
\frac{d}{dt}\, g(t,x) = \im \beta\(t,X_t(x)\)+\kappa\(t,X_t(x)\)\frac {|\tilde
\alpha_0(t,x)|^{2\sigma}}{|J_t(x)|^\sigma}, 
$$
with $g\big|_{t=0}=0$. Inserting $|\tilde \alpha_0(t,x)|=|a_I(x)|$ yields
the lemma, since $x\mapsto X_t(x)$ is a diffeomorphism of $\R^d$ for
fixed $t\in[0,\tau[$.
\end{proof}
\begin{remark}
Note that along the flow 
\begin{equation*}
\beta\(t,X_t(x)\)=\big <\chi_n\(\cdot, \nabla_x \phi
\(t,X_t(x)\)\),\, \frac{d}{dt}  
\chi_n\(\cdot, \nabla_x \phi \(t,X_t(x)\)\) \big>_{L^2(Y)},
\end{equation*}
which is exactly the same expression as given in \cite{GRT}, there however 
the authors do not distinguish between $a_0$ and $\tilde \alpha_0$. 
\end{remark}
So far we explicitly constructed an approximate solution, which solves  
\eqref{nls} up to terms of order $O(\e)$, since $u_1$ is not fully
defined yet. To obtain a better
approximation we need to set the term $b_2$ in \eqref{exp} equal to zero, which gives  
\begin{equation}
\label{neo}
\begin{aligned}
H_{\Gamma} (\nabla_x \phi) u_2+\left(U(x)+\partial_t \phi  \right)
 u_2= \ & L_1 u_1 + L_2 u_0 \, -\\
-\,  \lambda(t) \Big((2\sigma+1)&
 |u_0|^{2\sigma}u_1+2\sigma|u_0|^{2\sigma-2}u_0^2 \overline{u}_1\Big)  ,
\end{aligned}
\end{equation}
where for $u_0(t,x,y)=a_0(t,x)\chi_n(y,\nabla_x\phi)$ we define
\begin{equation*}
L_2 u_0 :=\frac{1}{2}\, \Delta_x u_0. 
\end{equation*}
Introduce the notations
\begin{equation}\label{eq:L_0+F}
L_0(t,x) = H_{\Gamma} (\nabla_x \phi)+U(x)+\partial_t \phi(t,x) \quad ;
\quad F(z)=|z|^{2\sigma}z\, .
\end{equation}
From \eqref{hamp}, $L_0$ is a $(t,x)$-dependent operator in $y$, and
since $\sigma\in \N$, $F$ is smooth. The following projector was used
to derive the transport equation \eqref{trans}:
\begin{equation}\label{eq:Pi_n}
\Pi_n(t,x)\( \sum_{j=1}^\infty
\alpha_j(t,x)\chi_j\(y,\nabla_x\phi(t,x)\)\) =
\alpha_n(t,x)\chi_n\(y,\nabla_x\phi(t,x)\) .
\end{equation}
Define $Q(t,x) = Id - \Pi_n(t,x)$. This operator is smooth, and a
partial inverse for $L_0$ can be defined on its range (by elliptic
inversion): $L_0^{-1}Q$ is 
well-defined, and smooth (up to caustics). 
Applying the operator $\Pi_n$ to \eqref{neo}, the
solvability condition reads 
\begin{equation}
\label{con1}
\int_{Y} \overline \chi_n(y,\nabla_x\phi) \, \(L_1 u_1 + L_2 u_0 
-
\lambda(t) \frac{d}{ds}F(u_0+su_1)\Big|_{s=0}\) 
\, dy =0.
\end{equation}
We decompose $u_1$ as
\begin{equation}
\label{u1}
u_1(t,x,y)=a_1(t,x)\chi_n\(y,\nabla_x \phi(t,x)\)+ u_1^\perp(t,x,y),
\end{equation}
where $a_1$  is some yet unknown function and $ u_1^\perp$ is such that
\begin{equation*}
\Pi_n(t,x) u_1^\perp(t,x,\cdot)  = \langle \chi_n(\cdot,\nabla_x
 \phi), u_1^\perp(t,x,\cdot)\rangle_{L^2(Y)}
 = 0, \quad  
\forall \, (t,x)\in [0,\tau[\times \R^d.
\end{equation*}
Now, $u^\perp_1$ is determined by \eqref{r01}:
\begin{equation}\label{eq:u_1perp}
u^\perp_1 = L_0^{-1}Q \(L_1 u_0 -\lambda(t)F(u_0)\)\, .
\end{equation}
which implies 
$u^\perp_1\in C^\infty([0,\tau[;\mathcal S(\R^d))$, since $u_0$ is, by
Lemma~\ref{le1}. Note that this relations \emph{imposes} a particular
form for the initial perturbation $\varphi_I^\e$, that is 
\begin{equation}\label{eq:corr1}
Q(0,x)\varphi_I^\e (x)= e^{i\frac{\phi_I(x)}{\e}}
\(L_0^{-1}Q\)(0,x) \(L_1 u_I -\lambda(0)F(u_I)\)
+\O(\e)\, .
\end{equation}
The term $\O(\e)$ will be defined more precisely later on. 
On the other hand, plugging \eqref{u1} into \eqref{con1} yields an
inhomogeneous \emph{linear} version of the transport equation
\eqref{trans} for $a_1$ (the 
propagating part of $u_1$): 
\begin{equation*}
\partial_t a_1 + \mathcal L a_1 - \beta(\nabla_x \phi, x)a_1+
i\lambda(t) \frac{d}{ds}F(u_0+s a_1)\Big|_{s=0}
=  {\tt S}(t,x),
\end{equation*}
where we may choose $a_1\big|_{t=0}=0$.
The complex-valued source term ${\tt S}(t,x)$ is given by
\begin{align}
{\tt S}(t,x) = i \Pi_n(t,x) \( L_1 u^\perp_1 + L_2 u_0 \)=
i\left< \chi_n(\cdot, \nabla_x\phi), \,   L_1 u^\perp_1 + L_2 u_0 
\right>_{L^2(Y)}. 
\end{align}
By this procedure, all higher order terms $u_j(t,x,y)$, $j\geq 1$, of
the asymptotic solution \eqref{wkb} can be obtained (recall that
the nonlinearity $F$ is smooth). 
Clearly we have that $u_j\in C^\infty([0,\tau[;\mathcal S(\R^d))$ for
all $j\geq 1$. At each step however, an additional condition must be
imposed recursively for the initial datum $\psi_I^\e$. This approach
is very similar to the one followed in \cite{DoRa}, except that the
Fourier modes are replaced by ``Bloch modes''. 
\newpar
Under the assumption \eqref{assi}, \eqref{ass}, we construct an
approximate   
solution, which solves \eqref{nls} up to a remainder $ O(\e^\infty)$,
provided that the initial data are well-prepared.  
To state precisely this property, define, for $N\ge 0$, 
\begin{equation}\label{eq:v}
\v_N^\e (t,x) :=  \ v_N^\e\(t,x,\frac{x}{\e}\)e^{i\phi(t,x)/\e}
\equiv \( \sum_{j=0}^N \e^j
u_j\(t,x,\frac{x}{\e}\)\)e^{i\phi(t,x)/\e}\, .  
\end{equation}
We will use the following spaces, for $s\in \N$: let
\begin{equation*}
\|f^\e\|_{X^s_\e}:=
\sum_{|\alpha|+|\beta|\leq s}\left\|
x^\alpha (\e \d)^\beta f^\e\right\|_{L^2}\, .
\end{equation*}
We define $X^s_\e$ as:
\begin{equation*}
X^s_\e :=\left\{ f^\e \in L^2(\R^d)\ ;\ \sup_{0<\e\leq 1}\|f^\e\|_{X^s_\e}  <+\infty\right\}.
\end{equation*}
These spaces are reminiscent of the spaces $H^s_\e(\R^d)$
introduced in \cite{Gues93} (see also \cite{RauchUtah}). There the 
dependence upon $\e$ is to recall that exactly one negative
power of $\e$ appears every time the approximate wave--function is
differentiated.  
In our case, such negative powers also appear because of the variable
$y$ and the substitution $y=x/\e$. The control of the momenta
is needed because of the potential $U$ (it would not
be needed in the proof of Theorem~\ref{theo:stab} below with $U$
sub-linear).  We can 
now state precisely the assumptions on the initial data:  
\begin{assumption}[Well-prepared initial data]\label{ass:varphi}
The initial data $\psi_I^\e$ satisfy Assumptions~\ref{assi} and
\ref{ass}, and for some $K\in \N$, the perturbation $\varphi_I^\e$
is of the form 
\begin{equation}\label{eq:K}
\varphi_I^\e(x)= e^{i\phi_I(x)/\e}
\sum_{j=1}^K \e^{j-1} \varphi_j(x,y)\Big|_{y=x/\e}
+\O\(\e^K\)\, , 
\end{equation}
where the $\O\(\e^K\)$ holds in $X^s_\e$ for any $s\in \N$. The
function $e^{i\phi_I/\e}\varphi_1$ is given by the first 
term of the right-hand side of \eqref{eq:corr1}, and if we denote
$\varphi_0 = u_I$, $\varphi_j(x,y)$ is given  recursively for $0\le
j\le K-2$ by 
\begin{equation*}
\varphi_{j+2} = \( L_0^{-1}Q\)(0,x) \( L_1 \varphi_{j+1}  +L_2
\varphi_j -\lambda(0) \frac{d^{j+1}}{ds^{j+1}} F \Big(u_I + \sum_{\ell
=1}^{j+1} s^\ell \varphi_\ell\Big)\Big|_{s=0}\)\, .
\end{equation*}
In the case $K=0$, the sum in \eqref{eq:K} is zero. 
\end{assumption}
\begin{remark}
We chose to impose $\Pi_n(0,x) \varphi_j(x,\cdot)=0$ for $j\ge 1$
(when we picked $a_1\big|_{t=0}=0$ for instance). Our approach
would also work with non-zero, smooth data $(\varphi_j)_{1\le j\le K}$
not necessarily 
satisfying this polarization property. All this approach is very
similar to the one followed in \cite{JolyRauch} to justify
nonlinear geometric optics for hyperbolic equations (see also
\cite{RauchUtah}, and \cite{DoRa} for the dispersive case). 
\end{remark}

We have the following Borel type lemma (see e.g. \cite{RauchUtah}):
\begin{lemma}\label{lem:Borel}
There exists $\widetilde\psi_I^\e \in \mathcal S(\R^d)$ satisfying
Assumption~\ref{ass:varphi}, such that \eqref{eq:K} holds for any
$K\in \N$.  
\end{lemma}
First, we will justify the asymptotics when the initial datum is given
by the above lemma. We will then show how to relax this assumption. Note that
the above approach is a nonlinear analog to the procedure followed in
\cite{PST}. In \cite{PST}, the authors construct $\e$-dependent
``super-adiabatic'' subspaces, in order to prove higher order
asymptotics in the linear case. In the present context, high order
asymptotics are needed to control the nonlinear terms (see
the proof of Theorem~\ref{theo:stab}). 
\begin{proposition}\label{prop:DA}
Let $\widetilde\psi^\e_I$ as in Lemma~\ref{lem:Borel}.  
Let $\tau >0$ be the time at which the first caustic is formed (if
any). Then for any $N\in \N$, $\v_N^\e$ solves
\begin{equation}
\label{approx}
\left \{
\begin{aligned}
i\e \partial _t \v_N^\e- H^\e \v^\e_N  = & 
\ \e \lambda(t)\, |\v_N^\e|^{2\sigma} \v_N^\e + \e^{N+1}r_N^\e,\\
\v_N^\e \big |_{t=0}   = &  \ \widetilde\psi^\e_I + \e^{N+1}\rho^\e_N,
\end{aligned}
\right.
\end{equation}
where $H^\e$ is defined by \eqref{ham} and $r_N^\e \in C^\infty(
[0,\tau[;\, \mathcal S(\R^d))$, $\rho^\e_N \in \mathcal S(\R^d)$
are such that $r_N^\e \in L^\infty_{\rm loc}([0,\tau[; X^s_\e)$ and 
$\|\rho^\e_N \|_{X^s_\e}=\O(1)$ for any $s\in \N$. 
\end{proposition}


\section{Nonlinear stability of the approximate
solution}\label{sec:stab} 

To prove that the above WKB--method yields a good approximation of the
exact solution, a nonlinear stability result is needed. First, we make
our assumptions on the potentials precise, and establish an 
existence result for \eqref{nls}. Next, we prove the validity of the
approximation derived above. 

\begin{assumption}\label{hyp:general} 
The potentials are smooth, real-valued: $V_\Gamma,U\in
C^\infty(\R^d;\R)$. 
\begin{itemize}
\item[(i)] $V_\Gamma$ is $\Gamma$-periodic, \ie it satisfies \eqref{eq:Vper}.
\item[(ii)] $U$ is  sub-quadratic:
$\d^\alpha U \in L^\infty(\R^d)\, ,\quad \forall \alpha\in \N^d \text{
such that }|\alpha|\geq 2$. 
\end{itemize}
\end{assumption}
\begin{remark}
The assumptions on $U$ include the cases of an isotropic harmonic
potential ($U(x)=|x|^2$), and of an anisotropic harmonic potential
($U(x)=\sum \omega_j^2 x_j^2$). It may also be taken equal to zero, or
incorporate a linear component $E\cdot x$, modeling a constant electric field
(\emph{Stark effect}, see \eg \ \cite{Cycon}).  
\end{remark}

\subsection{Existence of solutions to \eqref{nls}}

\begin{lemma}\label{lem:existence}
Let Assumption~\ref{hyp:general} be satisfied, and let $\psi^\e_I\in
\S(\R^d)$, the Schwartz space. Let $s>d/2$. Then there exists $t^\e>0$
and a unique  $\psi^\e \in C(]-t^\e,t^\e[;H^s(\R^d))$  solution to
\eqref{nls}. Moreover, $x^\alpha\psi^\e \in 
C(]-t^\e,t^\e[;H^s(\R^d))$ for any $\alpha\in\N^d$, $s\in \N$, and the
following conservation holds:
\begin{equation*}
\frac{d}{dt}\|\psi^\e(t)\|_{L^2} =0\, .
\end{equation*}
\end{lemma}
\begin{proof}
Since the dependence upon $\e$ is irrelevant at this stage, the above
statement follows from the study of
\begin{equation}
\label{nlsgen}
i \partial _t \psi =  -\frac{1}{2}\Delta \psi +
W(x)\psi + \lambda(t)\, |\psi|^{2\sigma} \psi\quad ;\quad
\psi \big |_{t=0}   =   \ \psi_I(x),\qquad\text{where:}
\end{equation}
\begin{itemize}
\item The potential $W$ is smooth, real-valued and sub-quadratic.
\item $\lambda(t)$ is a smooth real-valued function.
\item $\sigma \in \N$. 
\item $\psi_I\in \S(\R^d)$. 
\end{itemize}
The dependence of the local existence time $t^\e$ upon
$\e$ appears with scaling.
Notice that the nonlinearity $z\mapsto
|z|^{2\sigma}z$ is smooth, because $\sigma \in \N$. Since $W$ is
sub-quadratic, the Hamiltonian 
$-\frac{1}{2}\Delta + W$ is essentially self-adjoint on
$C^\infty_0(\R^d)$ (see for instance \cite{ReedSimon2}). The
assumption $s>d/2$ yields $H^s(\R^d)\subset
L^\infty(\R^d)$. Therefore, local existence and uniqueness in
$H^s(\R^d)$ follow from a fixed point argument, using Schauder's
lemma (see \eg \ \cite{Caz,RauchUtah}). 

To prove higher order regularity of $\psi$ and its momenta, one can
follow the proof of \cite{HNT} (see also \cite{Caz}). That article is
for the case $W\equiv 0$; the proof uses
Strichartz inequalities, following from dispersion estimates. When $W$
is smooth, real-valued and sub-quadratic, the same dispersion
estimates are available (\cite{Fujiwara79,Fujiwara}), and they imply
the same Strichartz inequalities (\cite{KT}).
Another difference with \cite{HNT} is that the Galilean operator
$x+it\nabla_x$ commutes with $i\d_t +\frac{1}{2}\Delta$, but in
general not with $i\d_t +\frac{1}{2}\Delta - W$. This is not a problem
in view of the above result, since
\begin{equation*}
\left[ x+it\nabla_x,W\right]= it\nabla W = \O\(1+|x|\) \, .
\end{equation*}
Thus, $\psi$, $x\psi$ and $\nabla_x \psi$ solve a coupled,
closed system of 
Schr\"odinger equations. A similar argument allows to treat higher
order momenta and derivatives. 

The conservation of the $L^2$-norm  follows from standard
arguments (see \cite{Caz}). 
\end{proof}

\begin{remark}
One cannot expect global existence in general. For instance, if
$\lambda(t)$ is a negative constant and if $\sigma >2/d$, finite time
blow-up may occur (see \eg \ \cite{Caz}). On the other hand, we shall
prove below that the solution $\psi^\e$ cannot blow-up before a
caustic is formed, at least for $\e$ sufficiently small.
\end{remark}
\textbf{Notation.} Let $(\alpha^\e)_{0<\e\leq 1}$ and $(\beta
^\e)_{0<\e\leq 1}$ be  
two families of positive numbers. In the following we shall write
\begin{equation*}
\alpha^\e \lesssim \beta^\e,
\end{equation*}
if there exists a $C>0$, independent of $\e \in ]0,1]$, such that 
\begin{equation*}
\alpha^\e \leq C \beta^\e,\quad \text{for all } \e \in ]0,1].
\end{equation*}
(The $C$ may very well depend on other parameters).

\subsection{Accuracy of the approximation} 
The main result we shall prove is the following:
\begin{theorem}[Stability result]\label{theo:stab}
Let $\psi^\e_I=\widetilde \psi^\e_I $ as in Lemma~\ref{lem:Borel}. Let
$\tau>0$  
given by \eqref{eq:caustique}, and 
$\v_N^\e$ given by \eqref{eq:v}. Then for any $\tau_0 \in ]0,\tau[$, 
there exists $\e_0>0$ such that for $0<\e\leq
\e_0$, the solution $\psi^\e$ to \eqref{nls} is defined up to time
$\tau_0$.
Moreover, for any $N\in \N$ and $s\in \N$, 
\begin{equation}\label{eq:O}
\sup_{0\leq t\leq \tau_0}\left\|\psi^\e(t) - \v_N^\e(t)\right\|_{X^s_\e} 
= \O\(\e^{N+1}\).
\end{equation} 
\end{theorem}
\begin{proof}
For $N\in \N$, we define the error term as $\w^\e_N := \psi^\e -
\v^\e_N$. From \eqref{nls} and \eqref{approx}, it solves
\begin{equation}\label{eq:w}
\left\{
\begin{aligned}
i\e\d_t \w^\e_N &= H^\e\w^\e_N  + \e
\lambda(t)\(|\psi^\e|^{2\sigma}\psi^\e - |\v_N^\e|^{2\sigma}\v_N^\e\)
- \e^{N+1}r^\e_N\, , \\
\w^\e_N\big|_{t=0} &=  \e^{N+1}\rho^\e_N\, ,
\end{aligned}
\right.
\end{equation}
where $H^\e$ is defined by \eqref{ham}. We start with the standard
energy estimate for Schr\"odinger equations: multiply the above
equation by $\overline{\w^\e_N}$, integrate over $\R^d$ and take the
imaginary part. Since $H^\e$ is self-adjoint, this yields
\begin{equation*}
\e \d_t \left\| \w^\e_N (t)\right\|_{L^2} \lesssim 
\e |\lambda(t)|\left\| |\psi^\e|^{2\sigma}\psi^\e -
|\v_N^\e|^{2\sigma}\v_N^\e \right\|_{L^2} +
\e^{N+1}\left\|r^\e_N(t)\right\|_{L^2} \, .
\end{equation*}
Since we work on the fixed, finite interval $t\in [0,\tau_0]$, the
smooth function $\lambda$ is bounded, and the above estimate implies:
\begin{equation}\label{eq:nrj1}
 \d_t \left\| \w^\e_N (t)\right\|_{L^2} \lesssim 
\left\| |\psi^\e|^{2\sigma}\psi^\e -
|\v_N^\e|^{2\sigma}\v_N^\e \right\|_{L^2} +
\e^{N}\left\|r^\e_N(t)\right\|_{L^2} \, .
\end{equation}
The idea is now to factor out $\w^\e_N$ in the right hand side of the
above inequality, and take advantage of the smallness of the source
term. To carry out this argument, we follow the method used to justify
(nonlinear) geometric optics for hyperbolic systems; we refer to
\cite{RauchUtah} for an expository presentation. 

Following \cite[Lemma~8.1]{RauchUtah} we have the following Moser-type lemma:

\begin{lemma}\label{lem:moser}
Let $R>0$, $s\in \N$, and $F(z)=|z|^{2\sigma}z$ for
$\sigma\in\N$. Then there exists $C=C(R,s,\sigma,d)$ such that if $\v$
satisfies
\begin{equation*}
\left\| x^\alpha (\e\d)^\beta \v \right\|_{L^\infty(\R^d)} \leq R
\quad \text{for all }|\alpha| +|\beta|\leq s\, ,
\end{equation*}
and $\w$ satisfies $\displaystyle \left\| \w \right\|_{L^\infty(\R^d)} \leq
R$, then 
\begin{equation*}
\sum_{|\alpha| +|\beta|\leq s}\left\| x^\alpha (\e\d)^\beta \(F(\v+\w)
- F(\v)\)\right\|_{L^2(\R^d)} 
\leq C \sum_{|\alpha| +|\beta|\leq s}\left\| x^\alpha (\e\d)^\beta
\w\right\|_{L^2(\R^d)}\, . 
\end{equation*}
\end{lemma}
\begin{proof}[Sketch of the proof of Lemma~\ref{lem:moser}]
When $X^k_\e$ is replaced by $H^k_\e$ (remove the control of the
momenta), the result is exactly \cite[Lemma~8.1]{RauchUtah}. The idea is to factor out 
$\w$ in the quantity $F(\v+\w) - F(\v)$ using the fundamental theorem of
calculus, then to use Leibniz' rule, to conclude with
Gagliardo--Nirenberg inequalities. In the case of $X^k_\e$, the
control of the momenta follows easily. 
\end{proof}

We first notice that $\v_N^\e$ is uniformly bounded in
$L^\infty([0,\tau_0]\times\R^d)$. To prove that $\w_N^\e$ is bounded
in $L^\infty([0,\tau_0]\times\R^d)$, we use a continuity argument, and
prove that it is actually small in that space, for $N$ sufficiently
large. This will be a consequence of the Gagliardo--Nirenberg
inequalities: 
\begin{equation}\label{eq:GN}
\text{for }s>d/2,\qquad
\|\w\|_{L^\infty(\R^d)} \lesssim \|\w\|_{H^s(\R^d)} \lesssim \e^{-d/2}
\|\w\|_{X^s_\e}\, .
\end{equation}
(The scaling factor $\e^{-d/2}$ is obvious when one uses Fourier
transform.)\\ 
By construction, $\w^\e_N(0,x)=\O\(\e^{N+1}\)$ in any space
$X^s_\e$. We first prove the result for $N$ sufficiently 
large, then show how to get rid of this assumption. From
Lemma~\ref{lem:existence}, there exists $t(\e,R)>0$ such that if
$N+1>d/2$, then for $\e$
sufficiently small, 
\begin{equation}\label{eq:solong}
\left\|\w_N^\e(t)\right\|_{L^\infty(\R^d)}\leq R
\end{equation}
for $t\in [0,t(\e,R)]$. As long as
\eqref{eq:solong} holds, 
\eqref{eq:nrj1} and Lemma~\ref{lem:moser} with $s=0$ imply
\begin{equation*}
 \d_t \left\| \w^\e_N (t)\right\|_{L^2} \leq C 
\left\| \w^\e_N (t) \right\|_{L^2} +
C\e^{N}\left\|r^\e_N(t)\right\|_{L^2} \, , 
\end{equation*}
and from Gronwall lemma, as long as \eqref{eq:solong} holds for $t\leq
\tau_0$, we get that
\begin{equation}\label{eq:L2}
\left\| \w^\e_N (t)\right\|_{L^2} \leq C \e^{N}\, .
\end{equation}
The idea is now to obtain similar estimates for the momenta and
derivatives of $\w^\e_N$. 

Applying the operator $\e\nabla_x$ to \eqref{eq:w} yields:
\begin{equation*}
\begin{aligned}
i\e\d_t (\e\nabla_x \w^\e_N) =&\,  H^\e(\e\nabla_x\w^\e_N)  + \e
\lambda(t)(\e\nabla_x )\(F(\psi^\e) - F(\v_N^\e)\)\\
&+ \left[ \e\nabla,H^\e\right]\w^\e_N - \e^{N+1}\e\nabla_x r^\e_N.
\end{aligned}
\end{equation*}
The same energy estimate as before gives:
\begin{align*}
\d_t \left\|\e\nabla_x \w^\e_N(t)\right\|_{L^2} \lesssim& \,
\left\|\e\nabla_x  
\(F(\psi^\e) - F(\v_N^\e) \)\right\|_{L^2} + \frac{1}{\e}\left\| \left[
\e\nabla,H^\e\right]\w^\e_N \right\|_{L^2} \\
&+ \e^{N}\left\|\e\nabla_x
r^\e_N\right\|_{L^2}.  
\end{align*}
But we have
\begin{equation*}
\left[\e\nabla,H^\e\right]  = \(\nabla V_\Gamma\)\(\frac{x}{\e}\) + \e
\nabla U(x)\, .
\end{equation*}
Since $\nabla V_\Gamma$ is bounded and $\nabla U$ is
sub-linear, the above estimate yields
\begin{equation}\label{H1}
\begin{aligned}
\d_t \left\|\e\nabla_x \w^\e_N(t)\right\|_{L^2} \lesssim &\,  \left\|
\e\nabla_x\(F(\psi^\e) - F(\v_N^\e)\) \right\|_{L^2} + \frac{1}{\e}\left\|
\w^\e_N \right\|_{L^2} + \left\|x
\w^\e_N \right\|_{L^2}\\
&+ \e^{N}\left\|\e\nabla_x 
r^\e_N\right\|_{L^2}\\
\lesssim &\,  \left\|\e\nabla_x\w^\e_N
 \right\|_{L^2} + \left\|x
\w^\e_N \right\|_{L^2}+ \e^{N-1}\,,
\end{aligned} 
\end{equation}
where we have used Proposition~\ref{prop:DA}, Lemma~\ref{lem:moser}
with $s=1$, and \eqref{eq:L2}. We see that when $U$ is
quadratic, we have to find a similar estimate for
$\|x\w^\e_N\|_{L^2}$. For that, multiply \eqref{eq:w} by $x$:
\begin{equation*}
i\e\d_t (x \w^\e_N) =  H^\e(x\w^\e_N)  + \e
\lambda(t)x\(F(\psi^\e) - F(\v_N^\e)\)
+ \left[x,H^\e\right]\w^\e_N - \e^{N+1}x r^\e_N.
\end{equation*}
Since $\left[x,H^\e\right]= -\e^2\nabla_x$, the energy estimate
yields, as long as \eqref{eq:solong} holds:
\begin{equation}\label{eq:x}
\begin{aligned}
\d_t \left\|x \w^\e_N(t)\right\|_{L^2} &\lesssim  \left\|x
\(F(\psi^\e) - F(\v_N^\e)\) \right\|_{L^2} + \left\| \e\nabla_x
\w^\e_N \right\|_{L^2} + \e^{N}\left\|\e\nabla_x 
r^\e_N\right\|_{L^2}\\
&\lesssim \left\|x \w^\e_N(t)\right\|_{L^2} + \left\| \e\nabla_x
\w^\e_N \right\|_{L^2} + \e^{N}\, .
\end{aligned} 
\end{equation}
Putting \eqref{H1} and \eqref{eq:x} together, we have:
\begin{equation*}
\d_t \(\left\| \e\nabla_x
\w^\e_N \right\|_{L^2}+ \left\|x \w^\e_N(t)\right\|_{L^2}\) \lesssim  
\left\| \e\nabla_x
\w^\e_N \right\|_{L^2} +\left\|x \w^\e_N(t)\right\|_{L^2}  + \e^{N-1}\, ,
\end{equation*}
and a Gronwall lemma yields, as long as \eqref{eq:solong} holds:
\begin{equation}\label{X1}
\left\| \w^\e_N(t)\right\|_{X^1_\e}\lesssim \e^{N-1}\, .
\end{equation}
One can check by induction that for $k\geq 0$,
so long as \eqref{eq:solong} holds,
\begin{equation}\label{Xk}
\left\| \w^\e_N(t)\right\|_{X^s_\e}\lesssim \e^{N-s}\, .
\end{equation}
We now take advantage of the Gagliardo--Nirenberg inequality
\eqref{eq:GN}. For $s>d/2$ and as long as \eqref{eq:solong} holds, we get
\begin{equation*}
\left\| \w^\e_N(t)\right\|_{L^\infty(\R^d)}\lesssim \e^{-d/2}\left\|
\w^\e_N(t)\right\|_{X^s_\e}\lesssim\e^{N-s-d/2}\,.  
\end{equation*}
Thus, if $N-s-d/2>0$, a continuity argument shows that
\eqref{eq:solong} holds up to time $\tau_0$ provided that $\e$ is
sufficiently small. In particular, $\w^\e_N$, hence $\psi^\e$, is well
defined up to time $\tau_0$ for $0<\e\leq \e(\tau_0)$. 
To complete the proof of Theorem~\ref{theo:stab}, we have to prove
\eqref{eq:O}. Fix $s,N\in \N$; let $s_1\geq s$ such that 
$s_1>d/2$, and $N_1\geq s_1+N+1$. 
We infer from \eqref{Xk} that
\begin{equation*}
\sup_{0\leq t\leq \tau_0}
\left\| \w^\e_{N_1}(t)\right\|_{X^{s_1}_\e}\lesssim
\e^{N_1-s_1}\lesssim \e^{N+1}\, .
\end{equation*}
It is straightforward that since $N_1>N$, 
\begin{equation*}
\sup_{0\leq t\leq \tau_0}\left\| \v^\e_N(t)-
\v^\e_{N_1}(t)\right\|_{X^{s_1}_\e}\lesssim \e^{N+1}\, .
\end{equation*}
We deduce that \eqref{eq:O} holds for any $s,N\in\N$. 
\end{proof}

\begin{remark}
A slightly shorter argument is available in the case $d\leq 3$, for
which we have $H^2(\R^d)\subset L^\infty(\R^d)$, to
 prove Theorem~\ref{theo:stab} in the case $s=2$ only. The idea is to
get an $X^2_\e$-estimate and use \eqref{eq:GN} again. Following an
idea due initially to T.~Kato \cite{Kato87}, consider the time
derivative of the error $\w^\e_N$. One can prove that $\|\e\d_t
\w^\e_N(t)\|_{L^2} =\O(\e^N)$, as long as \eqref{eq:solong}
holds. Plugging this into \eqref{eq:w}, we have, from \eqref{eq:L2} and
since $V_\Gamma$ is bounded and $U$ is sub-quadratic:
\begin{equation*}
\left\|\e^2\Delta \w^\e_N(t)\right\|_{L^2}\lesssim \e^N +
\left\|x^2\w^\e_N(t)\right\|_{L^2} \, .
\end{equation*}
The control of $\|x^2\w^\e_N(t)\|_{L^2}$ is then similar to \eqref{eq:x}:
\begin{equation*}
\left\|x^2 \w^\e_N(t)\right\|_{L^2}\lesssim \e^N +
\left\|x^2\w^\e_N(t)\right\|_{L^2} + \left\|\e^2\Delta
\w^\e_N(t)\right\|_{L^2} \, ,
\end{equation*}
and we can conclude as above. 
\end{remark}

Now it is easy to deduce the estimate announced in
Theorem~\ref{theo:typique}, when $\psi_I^\e$ is as in
Lemma~\ref{lem:Borel}. The $L^2$ estimate is \eqref{eq:O} with 
$N=s=0$. We have an $L^\infty$
estimate, mimicking the above proof: for $s>d/2$ and $N-d/2\geq 1$,
\eqref{eq:O} and \eqref{eq:GN} yield
\begin{equation*}
\sup_{0\leq t\leq \tau_0}\left\|\psi^\e(t) -
\v_N^\e(t)\right\|_{L^\infty(\R^d)}  \lesssim \e^{-d/2}\sup_{0\leq
t\leq \tau_0}\left\|\psi^\e(t) - 
\v_N^\e(t)\right\|_{X^s_\e}\lesssim \e^{N-d/2}\lesssim \e\, .
\end{equation*}
It is straightforward that 
\begin{equation*}
\sup_{0\leq t\leq \tau_0}\left\|\v^\e_0(t) -
\v_N^\e(t)\right\|_{L^\infty(\R^d)}  \lesssim \e\, ,\quad \text{hence }
\sup_{0\leq t\leq \tau_0}\left\|\psi^\e(t) -
\v_0^\e(t)\right\|_{L^\infty(\R^d)}  \lesssim \e\, .
\end{equation*}
Finally, we remove the assumption that $\psi_I^\e$ is as in
Lemma~\ref{lem:Borel}.
\begin{proposition}\label{prop:finale}
Let $\widetilde \psi^\e$ be the solution to \eqref{nls} with initial
datum $\widetilde \psi_I^\e$ as in Lemma~\ref{lem:Borel}. Let
$\psi_I^\e$ satisfying Assumptions~\ref{ass}, \ref{assi} and
\ref{ass:varphi} with $K\ge d$, and let $\psi^\e$ be the solution to
\eqref{nls} with initial datum $\psi_I^\e$. Then for any $\tau_0 \in
]0,\tau[$, there exists $\e_0>0$ such that for $0<\e\leq
\e_0$, $\psi^\e$ is defined up to time
$\tau_0$.
Moreover,  
\begin{equation*}
\sup_{0\leq t\leq \tau_0}\left\|\psi^\e(t) - \widetilde
\psi^\e(t)\right\|_{X^s_\e}  
= \O\(\e^{K+1 -s}\)\, ,\quad \text{for }s \ge 0\, .
\end{equation*} 
\end{proposition}
\begin{remark}\label{rem:well}
We deduce that Theorem~\ref{theo:typique} holds
with an $\O\(\e^d\)$
corrector in the initial datum. The $L^2$ estimate in
Theorem~\ref{theo:typique} is 
straighforward, using Theorem~\ref{theo:stab}. The $L^\infty$ estimate
\eqref{eq:asymLinfty} follows the same way, from
\eqref{eq:GN}.  Notice that the larger $K$, the
more precise asymptotics we infer; for example, if $K>d$, we can remove the
restriction $\eta>0$ in \eqref{eq:asymLinfty}, using the above
estimates and \eqref{eq:GN}. When
$s>K+1$, the above estimate does not look so good, since from
Theorem~\ref{theo:stab}, $\widetilde\psi^\e$ is bounded in
$X^s_\e$. Yet, it gives some non-obvious control on $\psi^\e$. 
\end{remark}
\begin{proof}[Sketch of the proof of Proposition~\ref{prop:finale}]
The proof is very similar to that of Theorem~\ref{theo:stab}, so we
shall be brief. Introduce $\w^\e = \psi^\e - \widetilde \psi^\e$. It
solves
\begin{equation*}
\left\{
\begin{aligned}
i\e\d_t \w^\e &= H^\e\w^\e  + \e
\lambda(t)\(|\psi^\e|^{2\sigma}\psi^\e - |\widetilde
\psi^\e|^{2\sigma}\widetilde \psi^\e\)\, , \\
\w^\e\big|_{t=0} &=  \O\(\e^{K+1}\)\quad \text{in }X^s_\e\text{ for any
}s\in\N\, .
\end{aligned}
\right.
\end{equation*}
We can then follow the same lines as in the proof of
Theorem~\ref{theo:stab}: there is no source term ($r_N^\e$ has
disappeared), and the size of $\w^\e$ is determined by the size of its
initial datum. We have
\begin{equation*}
\| \w^\e_{\mid t=0} \|_{L^\infty(\R^d)} \lesssim \e^{-d/2} \|
\w^\e_{\mid t=0} \|_{X^s_\e}\lesssim \e^{K+1-d/2}\, , \quad
\text{provided that } s>\frac{d}{2}\, . 
\end{equation*}
Since $K+1>d/2$, we can start the argument of
Theorem~\ref{theo:stab}. Theorem~\ref{theo:stab} and Sobolev
inequalities provide all the estimates we need for the ``approximate''
solution $\widetilde \psi^\e$; resuming all the arguments yields, so
long as \eqref{eq:solong} holds, 
\begin{equation*}
 \| \w^\e(t) \|_{X^s_\e}\lesssim \e^{K+1-s}\, .
\end{equation*}
Note that even if $K+1-s<0$, we can apply a Gronwall argument to prove
the above estimate. Since $K+1>d$, we can choose $s>d/2$ (not
necessarily an integer, but this causes no trouble, by interpolation)
such that $K+1-s>d/2$. The above estimate and \eqref{eq:GN} show that
\eqref{eq:solong} 
holds up to time $\tau_0$, for $\e\ll 1$. 
\end{proof}


\section{Generalizations and consequences}\label{sec:disc}

\subsection{Eigenvalue with multiplicity}
As a first generalization we remark that all given results could be 
extended to the case where $E_n(k)$ is an \emph{isolated} but \emph{$m$-fold degenerate}
family of eigenvalues, \ie  
$$
E_n(k) = E_*(k),\quad \forall \, n\in I\subset \N,\, |I|=m.
$$ 
Under the assumption (see \eg \ \cite{Ne} for a discussion on this) 
that there exists a smooth orthonormal basis $\{\chi_l(k,y))\}_{l\in
I}$ of $\ran \Pi_I(k)$,  
where 
$$
\Pi_I(k):=\sum_{l=1}^m \left | \chi_l(k)\right > \left < \chi_l(k)\right| 
$$  
denotes the spectral projector corresponding to $E_*(k)$, the 
appropriate two-scale WKB--ansatz would then be 
\begin{equation}
\label{gan}
\psi^\e\left(t,x,\frac{x}{\e}\right) \sim \sum_{l=1}^m
\, a_{0,l} (t,x)
\chi_l\left(\frac{x}{\e},\nabla_x\phi(t,x)\right)e^{i\phi(t,x)/\e}+
\O(\varepsilon),  
\end{equation} 
with $\phi(t,x)$ given by the solution of the Hamilton-Jacobi equation (\ref{hj}) with 
$E_n(k)\equiv E_*(k)$. As in \cite{PST, Te} this would then lead
to \emph{matrix-valued} transport equations, which in our case are all
coupled through the nonlinear term. The analysis of this system is
analogous to the scalar case  but leads to rather intricate and
tedious computations, which is why we neglected this situation. Also,
from the physical point of view it is known that for periodic
potentials such degeneracies are rather exceptional. (For the study of
a similar $2$-fold degenerated situation we refer to \cite{SpMa},
where a semi-classical scaled nonlinear Dirac equation is analyzed.)  

\subsection{Wigner measures}

Since Theorem~\ref{theo:stab} yields strong asymptotics for the
wave--function in $L^2(\R^d)$, we 
can compute the \emph{Wigner measure} associated to the family
$(\psi^\e)_{0<\e\leq 1}$. The Wigner measure of a family
$\(\psi^\e(t,\cdot)\)_{0<\e\leq 1}$ bounded in $L^2(\R^d)$ is the weak
limit (up to the extraction of a subsequence) of its Wigner transform,
\begin{equation}\label{wig}
W^\e \left[\psi^\e(t)\right](x,\xi) = \int_{\R^d}
\psi^\e\left(t,x-\frac{\e}{2}\eta  
\right)\overline{\psi^\e} \left(t,x+\frac{\e}{2}\eta
\right)e^{i\xi \cdot \eta}\, \frac{d\eta}{(2\pi)^d}. 
\end{equation}
This limit is then found to be a nonnegative Radon measure on phase space. The 
Wigner transform has proved to be
an efficient tool in the study of semi-classical and homogenization
limits  (see \eg \ \cite{BFPR,Ge,GMMP,LionsPaul}).  
\begin{corollary} Let $\psi^\varepsilon(t)$ be the unique
local-time-solution of~\eqref{nls}  
on $[0, \tau_0]$, as guaranteed by Theorem~\ref{theo:stab}, and let
$W^\varepsilon [\psi^\varepsilon(t)]$ be its Wigner transform.
Then, up to extraction of subsequences, we have
\begin{align}
\lim_{\varepsilon \rightarrow 0} W^\varepsilon [ \psi^\varepsilon] = \mu
\quad\mbox{in $\mathcal S'([0,\tau_0)\times \mathbb R_x^d\times
\mathbb R_{\xi}^d)$ weak-$\star$,} 
\end{align}
where the Wigner measure $\mu(t)$ of $\psi^\e(t)$ is given by
\begin{align}\label{wigm}
\mu(t,x,\xi)= \frac{|a_I(x)|^2}{|J_t(x)|} dx\otimes\sum_{\gamma^*
 \in \Gamma^*}  
 \left|\int_{\mathbb T^d} \chi_n(y,k)e^{-i y \cdot \gamma^*  } \, 
\frac{dy}{ (2\pi)^{d}} \right|^2\delta
(\xi-k-\gamma^*), 
\end{align}
with $k=\nabla_x\phi(t,x)\in \mathcal B$.
\end{corollary}
\begin{proof} We have to compute 
$$
\lim_{\varepsilon \rightarrow 0}\int_{\R^{2d}} f(x,\xi) W^\e
\left[\psi^\e(t)\right](x,\xi)dx\, d\xi  
= \int_{\R^{2d}} f(x,\xi) \mu(t,dx,d\xi),
$$
for any smooth test-function (observable) $f\in \mathcal
S(\R^d_x\times\R^d_\xi)$. To this end, we  
plug the approximation $\v^\e_0$ into the left hand side of this
relation (that is, we use the strong $L^2$ convergence stated in
Theorem~\ref{theo:typique}). 
Since $\chi_n(y,k)$ is $\Gamma$-periodic w.r.t. 
$y\in \R^d$, we can rewrite it in form of a Fourier series:
$$
\chi_n(y,k)= \frac{1}{(2\pi)^d} \sum_{\gamma^*\in \Gamma^*} e^{i
y\cdot \gamma^*}\, 
\int_{\mathbb T^d} \chi_n(z,k)e^{-i z \cdot \gamma^*  } \, dz .
$$
Using this representation, a non-stationary phase argument shows
that all ``non-diagonal'' terms in~\eqref{wig}  
vanish in the limit $\e \rightarrow 0$ and hence~\eqref{wigm} is
obtained from a straightforward  
computation.
\end{proof}
In our case, the strong convergence stated in Theorem~\ref{theo:stab}
shows that the Wigner measure of $\(\psi^\e(t,\cdot)\)_{0<\e\leq 1}$ is
the same as in the linear case (see \cite[Sect.~5.1]{GMMP}), since the
main nonlinear effect   
appears as an order $\O(1)$ phase $\omega$, defined in
Theorem~\ref{theo:typique}.  
In other words, the Wigner measure  does not ``see'' the
nonlinearity. This can be 
compared with the Wigner 
measures studied in \cite{CaWigner}, for equations similar to
\eqref{nls}, without potential. For the same scaling as in
\eqref{nls}, the main nonlinear effect was a ``slowly'' varying phase,
which was invisible to the Wigner measure. It only appears as the
first order correction in the  Wigner transform. 

\subsection{Complex-valued coupling factor}

When the coupling factor $\lambda(t)$ is not real-valued, the analysis
may be completely different; the approximate
solution may blow up before the caustic. 
The first hint is that the $L^2$--norm of $\psi^\e$ is not formally
conserved. Multiply \eqref{nls} by $\overline{\psi^\e}$, integrate
over $\R^d$ and take the imaginary part:
\begin{equation*}
\frac{d}{dt}\left\|\psi^\e(t)\right\|_{L^2}^2 = 2\im \lambda(t)
\left\|\psi^\e(t)\right\|_{L^{2\sigma+2}}^{2\sigma+2}\, .   
\end{equation*}
On the other hand, the formal analysis
of Sections~\ref{sec:DA.1} and \ref{sec:DA.2} still yields the transport
equation \eqref{trans}, which can also be written as
\eqref{eq:transray}.  Multiply \eqref{eq:transray} by
$\overline{\tilde a_0}$
and take the real part:
\begin{equation*}
\begin{aligned}
\frac{d}{dt}|\tilde a_0(X_t)|^2 &= -\im \kappa(X_t) \frac{|\tilde
a_0(X_t)|^{2\sigma+2}}{ |J_t|^\sigma}\\
&\equiv \im\lambda(t)\frac{|\tilde
a_0(X_t)|^{2\sigma+2}}{ |J_t|^\sigma}\int_Y \left|
\chi_n(y,\nabla_x\phi)\right|^{2\sigma+2}dy \, .
\end{aligned}
\end{equation*}
The solution of this ordinary differential equation may blow up in
finite time before a caustic is formed, and the WKB--analysis breaks
down at blow-up time. The above equation for the evolution of
$\|\psi^\e(t)\|_{L^2}^2 $ suggests that the exact solution may also
blow up. In that case, the limitation for the validity of the
WKB--expansion would not be a drawback of the method (as it is in the
case of caustics), but a genuine nonlinear effect.


\appendix

\section{Derivation of the leading order transport
equation}\label{sec:appA} 

For the benefit of the reader, we shall discuss here in more detail how
to pass from \eqref{cond} to  \eqref{trans}.

\newpar
First, it will be convenient to rewrite \eqref{l1} in a more symmetric
form   
\begin{equation*}
L_1 u_0 = i \partial_t u_0 - \frac{1}{2} \left[D_x\cdot(D_y+ \nabla_x
\phi) +  (D_y + \nabla_x \phi)\cdot D_x\right] u_0,
\end{equation*}
where from now on $D_x:=-i\nabla_x$. Then, inserting 
$$
u_0(t,x,y)=a_0(t,x) \chi_n(y,\nabla_x\phi),
$$
and denoting
$$
g_n(t,x,y) = \chi_n\(y,\nabla_x\phi(t,x)\)  ,
$$
the solvability condition \eqref{cond} can be written as  
\begin{equation}
\label{aptr}
\begin{aligned} 
& \ \partial_t a_0 + \< g_n ,\, \d_t g_n  \>_{L^2(Y)}\, a_0 
+ \frac{1}{2}\< g_n  ,\, \nabla_x\cdot(D_y+ \nabla_x \phi)\, (a_0
g_n)\>_{L^2(Y)} \\  
& \ + \frac{1}{2}\< g_n , \, (D_y+ \nabla_x \phi)\cdot \nabla_x(a_0
g_n)  \>_{L^2(Y)}   
- i \kappa (t,x) |a_0|^{2\sigma} a_0 = 0.
\end{aligned}
\end{equation}
Here we have used definition \eqref{k} and the fact that $\left<
\chi_n,  \chi_n \right>_{L^2(Y)}= 1$.  
Differentiating the eigenvalue equation \eqref{bloch} w.r.t. to $k$ yields
\begin{equation}
\label{id0}
(\nabla_k H_{\Gamma}(k)- \nabla_k E_n(k))\chi_n+
(H_{\Gamma}(k)- E_n(k))\nabla_k \chi_n=0.
\end{equation}
Taking in this identity the scalar product with $\chi_n$ we obtain
\begin{equation}
\label{id1}
\begin{aligned}
\left< \chi_n , \,  \nabla_k H_{\Gamma}(k)\chi_n \right>_{L^2(Y)}
\equiv & \, \left<  \chi_n , \,  (D_y+k) \chi_n  \right>_{L^2(Y)} \\
= & \, \nabla_k E_n(k),
\end {aligned}
\end{equation}
since $H_\Gamma $ is self-adjoint. 
From \eqref{id1} we deduce that \eqref{aptr} can be written as 
\begin{equation}
\label{aptr1} 
\partial_t a_0 + \left<  g_n , \, \d_t g_n  \right>_{L^2(Y)}\, a_0 
+ \nabla_k E_n(\nabla_x \phi) \cdot \nabla_x a_0 + f(t,x)\, a_0 =  i
\kappa (t,x) |a_0|^{2\sigma} a_0, 
\end{equation}
where 
$$ 
f(t,x)= \frac{1}{2}\left<  g_n , \,  (D_y+\nabla_x \phi)\cdot
\nabla_x g_n  \right>_{L^2(Y)} + \frac{1}{2} 
\left<  g_n , \,   \nabla_x \cdot (D_y+\nabla_x \phi) g_n \right>_{L^2(Y)} .
$$
Next, we substitute $\chi_n$ by $g_n$ in \eqref{id1} and differentiate  w.r.t. $x\in \R^d$:  
\begin{equation*}
\left< \nabla_x g_n, \,  (D_y+\nabla_x\phi) g_n \right>_{L^2(Y)} +
\left<g_n ,\, \nabla_x \cdot (D_y+\nabla_x\phi) g_n \right>_{L^2(Y)}  =
\diverg_x\nabla_k E_n(\nabla_x\phi). 
\end{equation*}
Since $D_y$ is
self-adjoint and $\nabla_x\phi$ is real, we have
\begin{equation*}
\alpha := \<  g_n ,\,  (D_y +\nabla_x\phi)\cdot \nabla_x g_n \>_{L^2(Y)} =
\<   (D_y +\nabla_x\phi)g_n ,\, \nabla_x g_n \>_{L^2(Y)} ,
\end{equation*}
and we infer from above that
$$
\alpha +\Delta_x\phi +\overline\alpha = \diverg_x\nabla_k
E_n(\nabla_x\phi).
$$ 
Therefore 
\begin{equation}
\label{apr}
f(t,x)=  \alpha +\frac{1}{2}\Delta_x\phi = \re \alpha
+\frac{1}{2}\Delta_x\phi +i \im \alpha\\
= \frac{1}{2}\diverg_x\nabla_k E_n(\nabla_x\phi) 
+ i \im \alpha.
\end{equation}
We simplify the last term.
From \eqref{id0}, with $k=\nabla_x\phi$, we obtain
\begin{equation*}
\( \(D_y +\nabla_x\phi\) -\nabla_kE_n(\nabla_x\phi)\)g_n  + \(
H_\Gamma (\nabla_x \phi)-E_n(\nabla_x\phi)\)\nabla_k
\chi_n\(y,\nabla_x\phi\) =0.
\end{equation*}
Taking the $L^2(Y)$-scalar product by 
\begin{equation*}
\d_{x_j} g_n =  \sum_{l=1}^d \d^2_{x_j x_l}\phi
\d_{k_l}\chi_n\(y,\nabla_x\phi\)  
\end{equation*}
and taking the imaginary part, we have, since
$\left< \chi_n,  \nabla_x \chi_n \right>_{L^2(Y)}\in i\R$:
\begin{equation}
\label{id2}
\begin{aligned}
\im \alpha = &-i\nabla_k E_n(\nabla_x \phi)\cdot\left< g_n,  \nabla_x
g_n \right>_{L^2(Y)} \\
 & - \sum_{j=1}^d \im  \big< \(H_\Gamma(\nabla_x \phi) -E_n(\nabla_x
\phi)\)\, \d_{k_j} \chi_n , \,\sum_{l=1}^d \d^2_{x_j x_l}\phi\, 
\d_{k_l}\chi_n  \big>. 
\end{aligned}
\end{equation}
The last sum also reads:
\begin{align*}
\sum_{1\leq j,l\leq d}\d^2_{x_j x_l}\phi
\im \< \(H_\Gamma(\nabla_x \phi) -E_n(\nabla_x
\phi)\)\, \d_{k_j} \chi_n , \,
\d_{k_l}\chi_n\>.
\end{align*}
Since $H_\Gamma$ is self-adjoint, this term is zero. 
Hence, \eqref{aptr1} together with \eqref{apr} and \eqref{id2} give
the following equation  for the principal amplitude:
$$ 
\partial_t a_0 + \left<  g_n ,\, \partial_t  g_n \right>_{L^2(Y)}\, a_0 
+ \mathcal L a_0 
+\nabla_k E_n(\nabla_x \phi)\cdot \left< g_n   , \, \nabla_x g_n  \right>
a_0
=   i \kappa (t,x) |a_0|^{2\sigma} a_0,
$$
where $\mathcal L$ is defined as in \eqref{l}. Finally, using the
Hamilton-Jacobi equation   
\eqref{hj}, a straightforward calculation shows 
\begin{equation}
\left< g_n,\,  \partial_t g_n\right>_{L^2(Y)} +\nabla_k E_n(\nabla_x
\phi)\cdot \left<  g_n , \,\nabla_x g_n \right> = -\beta (t, x) 
\end{equation} 
and we conclude that $a_0$ satisfies the nonlinear transport equation
\eqref{trans}.


\noindent {\bf Acknowledgments}. The authors are grateful to Stefan
Teufel for fruitful discussions on this work.

\bibliographystyle{amsplain}
\bibliography{bloch}

\end{document}